\documentstyle[preprint,prb,aps]{revtex}
\begin{document}
\draft

\title{Universality at integer quantum Hall transitions
}

\author{Kun Yang, D. Shahar, R. N. Bhatt, D. C. Tsui and M. Shayegan}
\address{Department of Electrical Engineering, Princeton University,
Princeton, NJ 08544}
\date{\today}

\maketitle

\begin{abstract}
We report in this paper results of experimental and theoretical studies of
transitions between different integer quantum Hall phases, as well as
transition between the insulating phase and quantum Hall phases at high
magnetic fields.
We focus mainly on universal properties of the transitions.
We demonstrate that properly defined conductivity tensor is universal
at the transitions. We also present numerical results of a non-interacting
electron model, which suggest that
the Thouless conductance is universal at integer quantum Hall transitions,
just like the conductivity tensor. Finite temperature and system size effects
near the transition point are also studied.

\end{abstract}

\pacs{71.30.+h, 73.40.Hm}

\section{Introduction}
\label{intro}

Continuous (or second order) quantum phase transitions in many-electron
systems are of general 
interest to condensed matter physicists.\cite{sondhi}
Recently a class of such 
quantum phase transitions, namely the transitions between different
quantum Hall plateaus, and the transition between a quantum Hall phase and
an insulating phase at high magnetic field ($B$), 
have been under extensive experimental 
\cite{wei,koch,jiang,wang,hughes,alphenaar,HPWei:Current,Wong95,Pan:Scaling,shahar95,okamoto,shahar96,Shahar:QHtrans} and
theoretical
\cite{chalker,theory,huo,klz,huo93,lee93,wang94,ld,ludwig,macdonald,lee,lee96,wang96,cho,ando,tan,liu,yang,xie,sheng,sorensen,sachdev,sw2} study.
In the renormalization group (RG) language, continuous phase transitions
are controlled by RG fixed points, and many properties of the transition
depend only on which fixed point the transition is controlled by, or which
universality class it belongs to, and independent of microscopic details 
of the system. The best known examples of such universal properties are 
of course the critical exponents. In principle, quantities that are 
dimensionless are possibly universal at the critical point.
Of particular interest in the study of quantum Hall transitions
is the conductivity tensor at the critical point, which in two dimensions (2D)
can be expressed 
as dimensionless numbers times the fundamental unit of conductance, ${e^2\over
h}$. It has been suggested that they should be universal, both at
superfluid-insulator transition,\cite{fisher} 
and quantum Hall transitions.\cite{klz}
In the latter case, which is the focus of the present paper, this suggestion
has received support from both experimental\cite{shahar95} study at the
quantum Hall-insulator transition at high magnetic field
(or transition in the lowest Landau level), and numerical
works in the lowest Landau level\cite{huo93} and the network model.
\cite{wang96,cho}

In this paper we will present further experimental evidence that supports the
universality of conductivity tensor at integer quantum Hall transitions, 
in both the lowest Landau level and {\em higher} Landau levels; we also 
demonstrate that the transitions in different Landau levels are in the 
same universality class.
We will also present results of numerical studies of the
integer quantum Hall transitions, based on non-interacting electron
models; our results suggest that another 
dimensionless quantity, the Thouless conductance, a quantity that is closely
related to the longitudinal conductance of the system, is also universal
at integer quantum Hall transitions.

By definition, a quantum critical point is a critical point at zero temperature
$(T=0)$. Real experiments, however, are always done at finite temperatures, and
the critical point can only be reached asymptotically in the low temperature
limit of the experiments. We will also present data on the $T$ dependence of
various physical quantities near the critical point, and study
how the universal 
values of the resistivity (or conductivity) tensor 
are reached asymptotically. In a {\em noninteracting} system, the 
effect of a finite $T$ is to set a finite dephasing length, which effectively
divide an otherwise infinite system into incoherent finite pieces, and 
introduce finite size effects to the critical behavior. We have therefore
also studied the size dependence of the Thouless conductance, near the 
critical point.

The paper is organized in the following way.
In section \ref{experiment} we present our experimental results on studies
of the resistivity tensor, both at the high field QH-insulator {\em and}
QH-QH critical points. We study their behavior not only in the asymptotic
low temperature limit, but also the temperature dependence at higher 
temperature, and how the universal values are approached as temperature 
decreases. In section \ref{theory} we present results of our numerical 
study on a non-interacting electron model on a lattice, and demonstrate that
at the critical points, the properly defined Thouless conductance is a
universal number that is independent of the strength and type of random
potential, amount of mixing between different Landau levels (subbands), 
whether there is particle-hole symmetry,
etc. We also study the system size dependence of the Thouless conductance
at the critical points,
and demonstrate that the universal asymptotic value is reached at surprisingly 
small system sizes in the lowest Landau subband. 
In section \ref{summary} we summarize our findings by discussing 
the relations between our experimental 
and numerical results, as well as their relevance to
existing theoretical and 
experimental
works on this subject.

\section{Experiments}
\label{experiment}
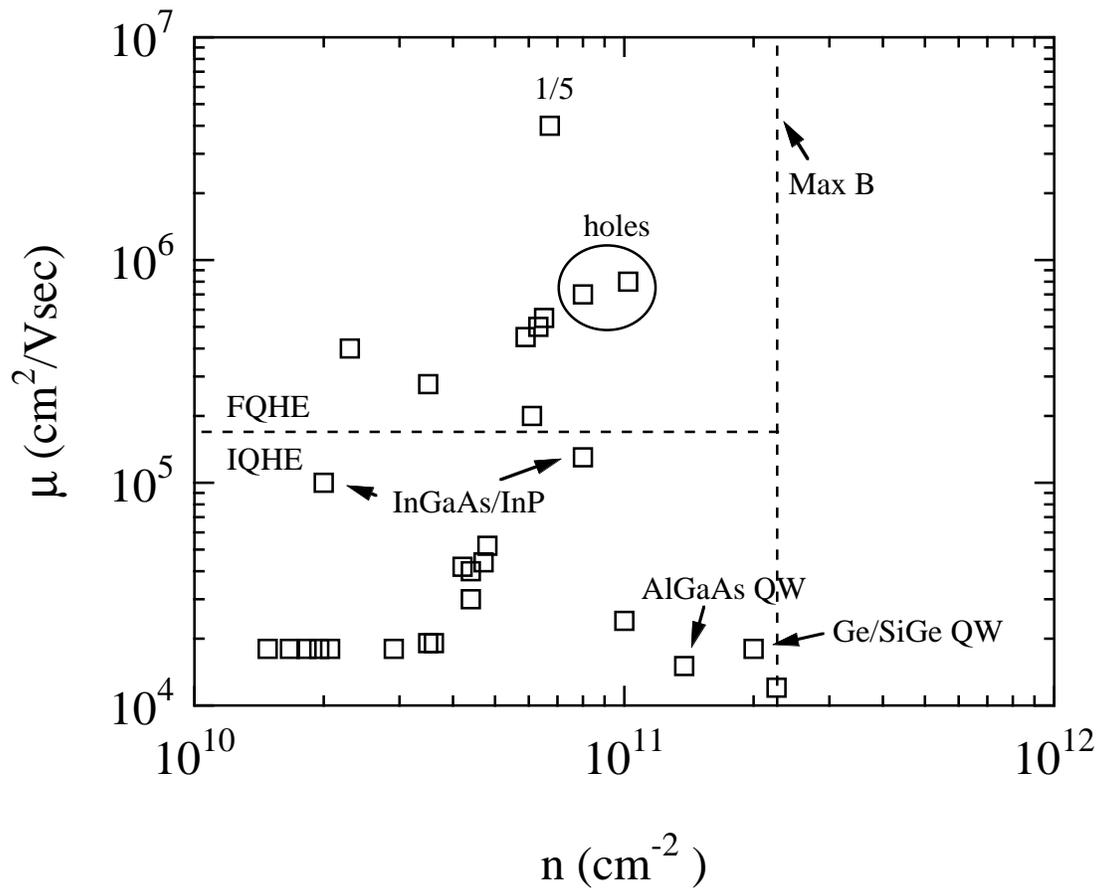
\begin{figure}
\caption{Mobility vs. density for some of the samples in this study. 
The vertical dashed line indicated by Max $B$ represents the maximum 
density for which the QH-insulator transition can be observed in our 
15.5 T magnet. The horizontal dashed line approximately seperates 
samples that do not exhibit the fractional QH from those that do. 
$p$-type samples are circled, and the sample labeled 1/5 exhibit the 
reentrant insulating transition which we do not discuss in this paper.
}
\label{mobilityDensity}
\end{figure}

In this section we describe our experimental results. While our main
finding has already been published before\cite{shahar95}, we will present here a more
detailed account of our study of the QH to Insulator at low $T$.
Since the main motivation of our work was to test for the
theoretically predicted {\em universal} features, we have set to study
a broad range of samples that represents much of the available 2DES
samples at the time of this work. In Fig. \ref{mobilityDensity} we
illustrate the large diversity of samples studied by plotting the
mobility ($\mu$) vs. density (n) of some of our samples. The
range of the axis in this figure is chosen to represent virtually
the entire range of 2DES that is reported in the literature
($\mu=10^{3}-10^{7}$ cm$^{2}$/Vsec and $n=10^{9}-10^{12}$ cm$^{-2}$).
As can be seen, our samples cover a significant area in of this
log-log graph. To increase the generality of our results we obtained
our samples from 6 different sources including 3 MBE, one LPE
and two MO-CVD machines. To insure that geometrical factors, which are
known to introduce modifications to transport in the QH regime, do not
come into play in our study, we did not maintain a uniform sample
geometry. Rather, our samples were cut in many different shapes: some
where wet-etched in a Hall-bar shape of various dimensions, and others
were cleaved in a square or rectangular shape with contacts diffused
along the edges. The smallest contact to contact dimension was
100 $\mu$m while the largest was 1 mm. Naturally, we did not adhere
to a specific structure design of the wafers and there too diversity
abound as we studied 2DES's embedded in GaAs/AlGaAs heterostructures
and quantum wells (QW's), InGaAs/InAlAs QW's, InGaAs/InP
heterostructures, AlAs/GaAs QW's, Ge/SiGe QW's and Si MOSFET's. The
total number of samples studied at low $T$'s exceeds 150.

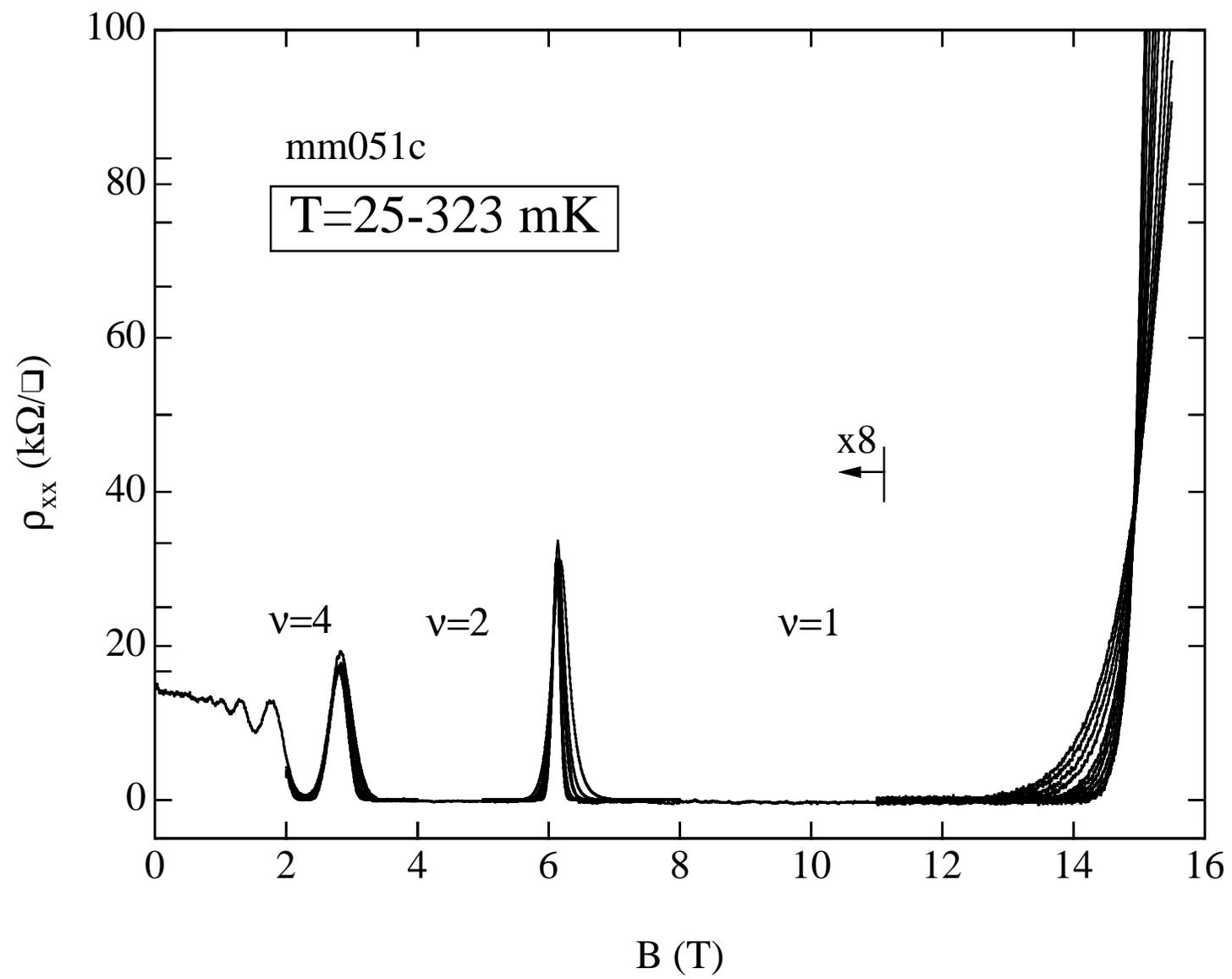
\begin{figure}
\caption{$B$ traces of $\rho_{xx}$ at $T=25$, 42, 62, 84, 106, 125, 
145, 194, 238, 284, 323 mK, for a GaAs/AlGaAs sample mm051c.
}
\label{TypTransition}
\end{figure}

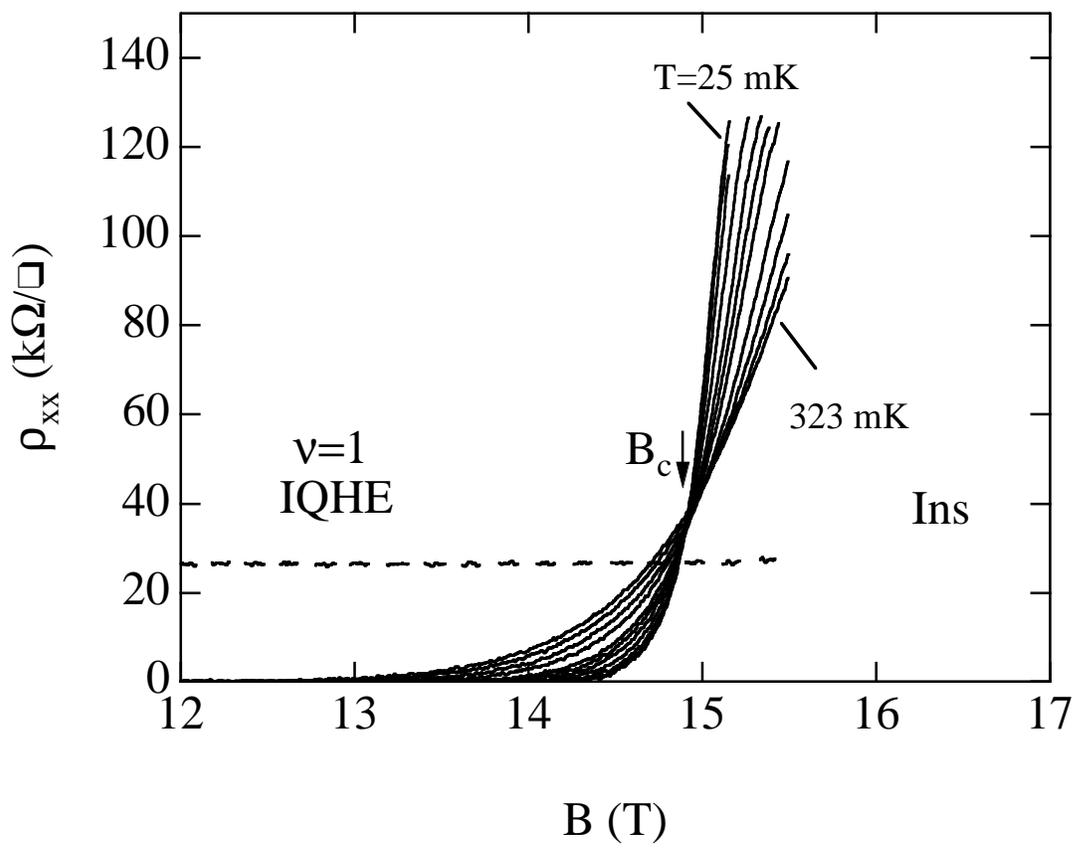
\begin{figure}
\caption{Same as Fig. \ref{TypTransition} for a narrow $B$ range 
focusing on the transition. We also included a $\rho_{xy}$ trace 
(dashed curve).
}
\label{CrossPoint}
\end{figure}

Quite generally, the QH series terminates, at high $B$, with a
transition to an insulating phase. A typical transition is shown in Fig.
\ref{TypTransition}, where we plot $B$ traces of the diagonal resistivity, $\rho_{xx}$,
taken at several $T$'s. The sample exhibit metallic behavior at low $B$
followed by a set of integer QH states manifested by minima in
$\rho_{xx}$. As $B$ is further increased beyond $\nu=1$, $\rho_{xx}$
increases for all $T$'s. If we examine, however, the $T$ dependence of
$\rho_{xx}$ focusing on the high-$B$ region (Fig. \ref{CrossPoint}), 
we observe  a stark change in its character: In the QH region, 
$\rho_{xx}$ increases with $T$, while
at higher $B$ the opposite occurs, and $\rho_{xx}$ decreases with $T$. It
is therefore reasonable to use the temperature coefficient of resistivity
(TCR) to delineate two different transport regimes: The QH state 
(TCR$<0$)
and the insulator (TCR$>0$). We adopt this empirical `definition' of the
phases for the rest of this paper.

Examining Fig. \ref{CrossPoint} further reveals another transport feature which is
typical to the QH-to-insulator transition. It is possible, at low $T$,
to identify a clear and well-defined $B$ value for which the TCR vanishes,
within experimental error. The existence of this critical $B$ value
($B_c$) allows us to unambiguously determine (subject to the definition of the
previous paragraph) the boundary between the QH
state and the insulator. A complete phase diagram can be obtained by
following the position of this `crossing' point as other relevant
parameters such as disorder or $n$ is changed, as was done extensively by
Wong {\em et al.}\cite{Wong95}, Song {\em et al.}\cite{song97} and others.

In a recent paper, Shahar {\em et al.}\cite{shahar95} reported on a study of the $\rho_{xx}$
value at $B_c$ ($\rho_{xxc}$) for a large collection of samples. They
noticed that,
in accordance with theoretical expectations\cite{huo93}, $\rho_{xxc}$ seems to be close to
the quantum unit or resistance, $h/e^2$, independent of sample parameters.
Further, they showed that this apparent universality holds also for
transitions from the $1/3$ fractional QH state to the insulator, again in
agreement with theoretical predictions\cite{klz}.

In the reminder of this
section we expand on these previous findings by concentrating
on the transport properties at the
critical point. We will provide further evidence for the notion of
universality near the QH-to-insulator transition, and remark on the QH-QH transitions.
We will also closely examine the transport at the critical point and
show that, at higher $T$'s,
systematic deviation of $\rho_{xx}$ are clearly seen, although their
trend and magnitude are sample dependent. Finally, we will discuss
the disorder-driven QH-insulator transition which is realized
experimentally by changing $n$ by means of a metallic 
front-gate\cite{jiang}. We will
emphasize the similarities between the disorder and magnetic field driven
transitions.

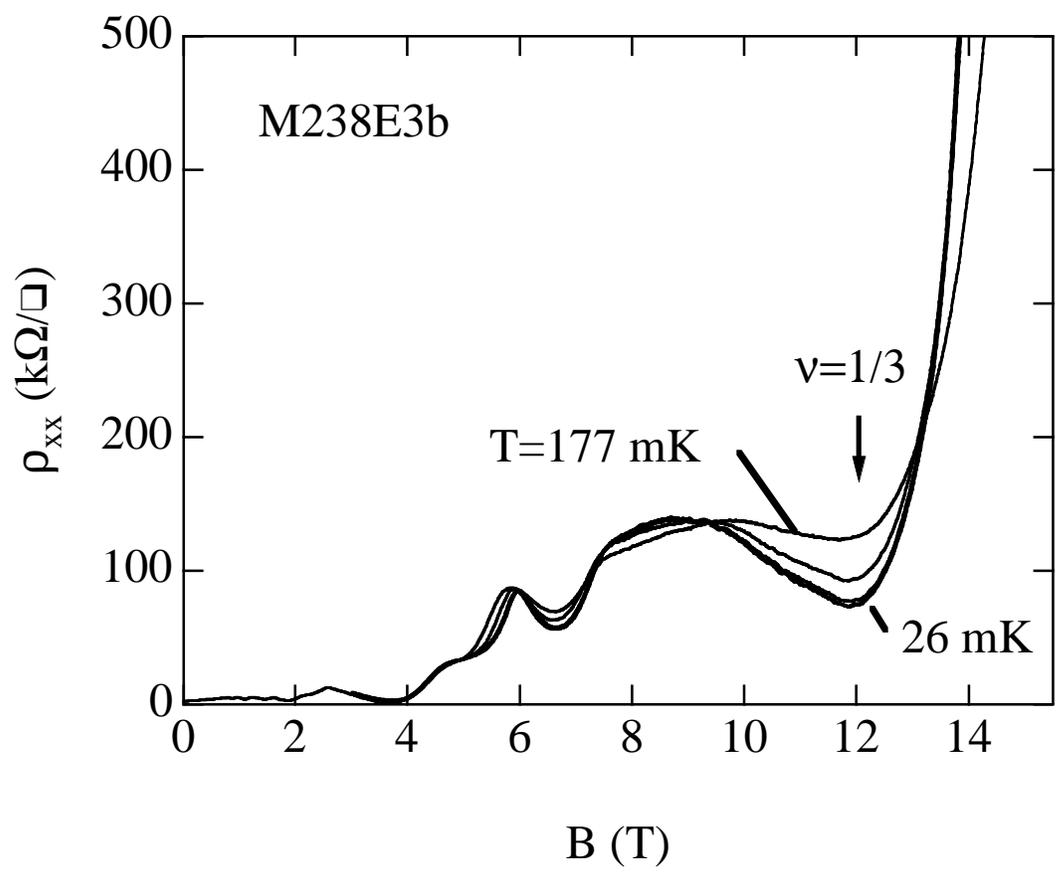
\begin{figure}
\caption{$B$ traces of $\rho_{xx}$ at $T=26$, 46, 55, 99 and 177 mK 
for a $p$-type GaAs/AlGaAs sample grown on (311)A substrate. Note the 
apparent crossing point at 220 k$\Omega$.}
\label{BadSample}
\end{figure}

We would like to precede the discussion of our data with a
note of caution. It would be an unreasonable
expectation that {\em all} samples will behave
uniformly in any subset of their transport characteristics. It is
well-known, for instance, that samples exhibiting the FQHE do not
undergo a transition from the $\nu=1$ IQHE to an insulator (they
will, ultimately, undergo a transition to an insulator at higher $B$
from a FQHE state). We therefore need to define clearly which
subset of our samples will be included in our test for universality at
the QH to insulator transitions. To clarify this need we plot, in
Fig. \ref{BadSample}, $B$ traces of $\rho_{xx}$ taken at several $T$'s
for a GaAs/AlGaAs sample grown on (311)A substrate to produce hole carriers. The ``QH''
state marked on the figure as $1/3$ is clearly abnormal with a very
high minimum resistivity at the lowest $T$. It is not surprising that
the following ``transition'' (common crossing point of the
different $T$ traces) to an insulator is at a very high
value, $\rho_{xxc}=220$ k$\Omega$, which significantly deviates from
universality. The above discussion leads to a natural, albeit arbitrary,
criterion for the ``suitability'' of a given transition for
a test of universality: For a transition to be considered, we require
that it exhibit a fully developed QH state followed by a strong insulating
behavior.\cite{RlawRemark} In both cases different
tests can be considered to define the ``strength'' of the phase, and we
chose for the QH a resistivity that decrease exponentially with
decreasing $T$ and, in addition, a value of $\rho_{xx}$ at the QH minimum
which is less than $h/100e^2$ at our lowest $T$, such that it is
undecernable from zero when plotted on a scale which includes the
transition point. Similarly we defined a fully developed insulator if
$\rho_{xx}$ increases monotonously with $B$ and $T$ and reaches a value
greater then $100h/e^2$ at our lower $T$. Arbitrary as they are, these
simple criteria safely eliminate from consideration samples like that 
of Fig. \ref{BadSample}.

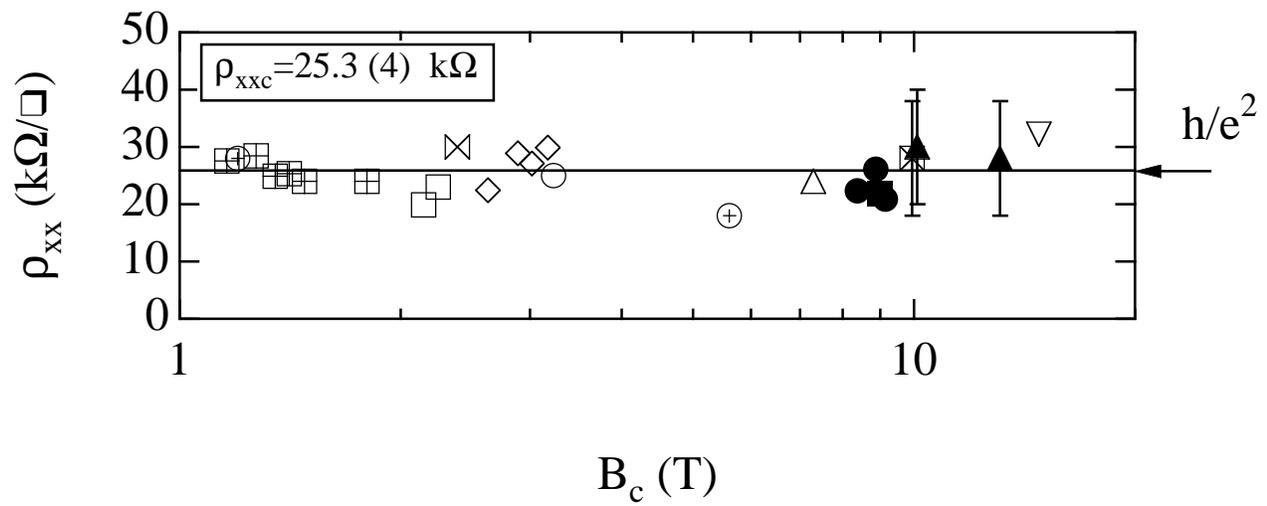
\begin{figure}
\caption{$\rho_{xx}$ vs. $B_{c}$ for some of our samples.
Empty (filled) symbols are  for transitions from the $\nu=1$ IQHE  
($\nu=1/3$ FQHE) state. The error bars are typically smaller than the 
symbol size, except for the samples where they are indicated, which 
had ill-defined geometries.
}
\label{RxxcBc}
\end{figure}

In a previous letter\cite{shahar95} we reported the observation of a universal value of
$\rho_{xxc}$. In Fig. \ref{RxxcBc} we plot $\rho_{xxc}$ vs. $B_c$ for 20 of our
samples. In fact, other absisca could have equally been used, such as
$\mu$ or $n$, as $\rho_{xxc}$ is rather independent of sample parameters
and appears to be scattered around $h/e^2$ (solid line in Fig. \ref{RxxcBc}). 
Two points should be emphasized. First, as we noted before, the $\rho_{xxc}$ value
is not significantly different between samples that undergo the
transition from the $\nu=1$ integer QH state (empty symbols) and those
for which the transition takes place from the $\nu=1/3$ fractional QH
state (solid symbols). This is in agreement with the theoretical notion
of super-universality\cite{klz}. We will not be discussing fractional QH
transitions in the remainder of this paper.

Second, the data points are scattered over a relatively wide range of
$\approx 25\%$ around $h/e^2$. This scatter is evident not only for
different samples but also for different cool-downs of the same sample.
This cool-down dependence of $\rho_{xxc}$ is particularly puzzling if we
recall that are samples are all of rather large size. This point has been
emphasized in our previous publication, and is not yet understood
theoretically.

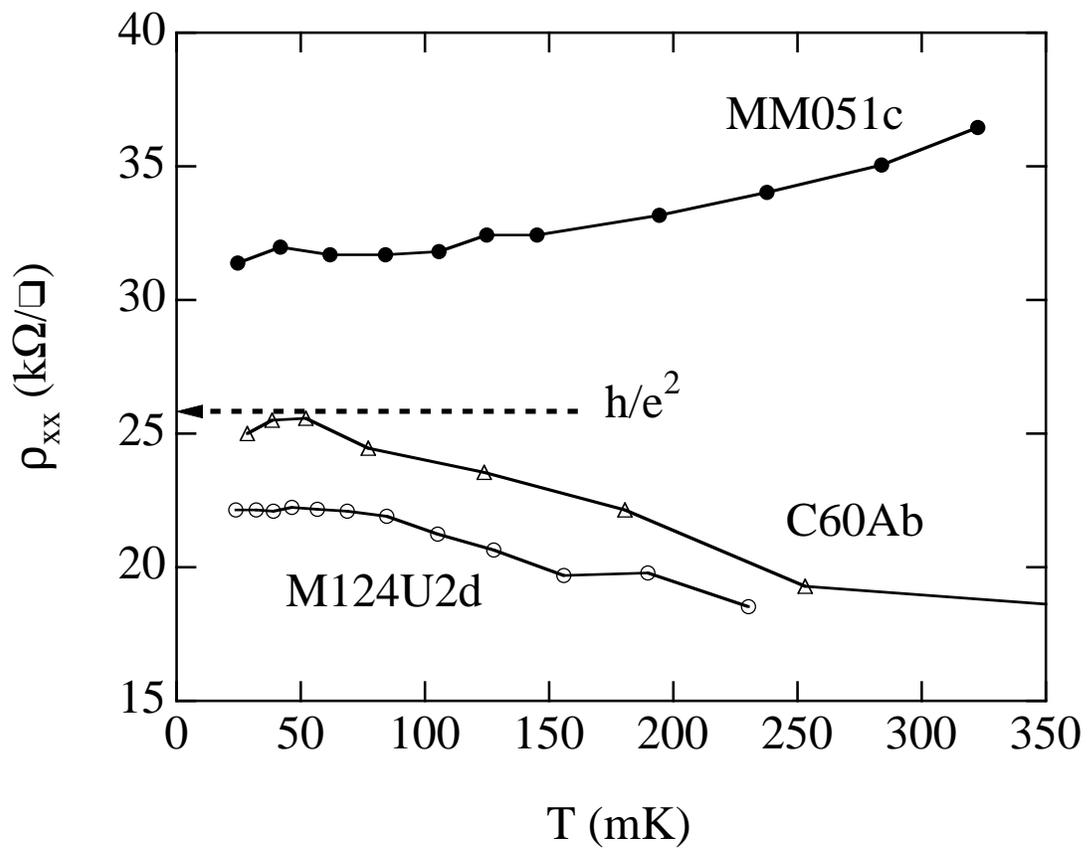
\begin{figure}
\caption{$\rho_{xx}$ at $B_{c}$, $\rho_{xx}$, vs. $T$ for three of our samples. The dashed arrow 
indicates the theoretically predicted value for the transition, $h/e^{2}$. 
Sample m124u2d exhibit the fractional QH effect, and the data 
depicted is from the $\nu=1/3$-insulator transition.
}
\label{RxxT}
\end{figure}

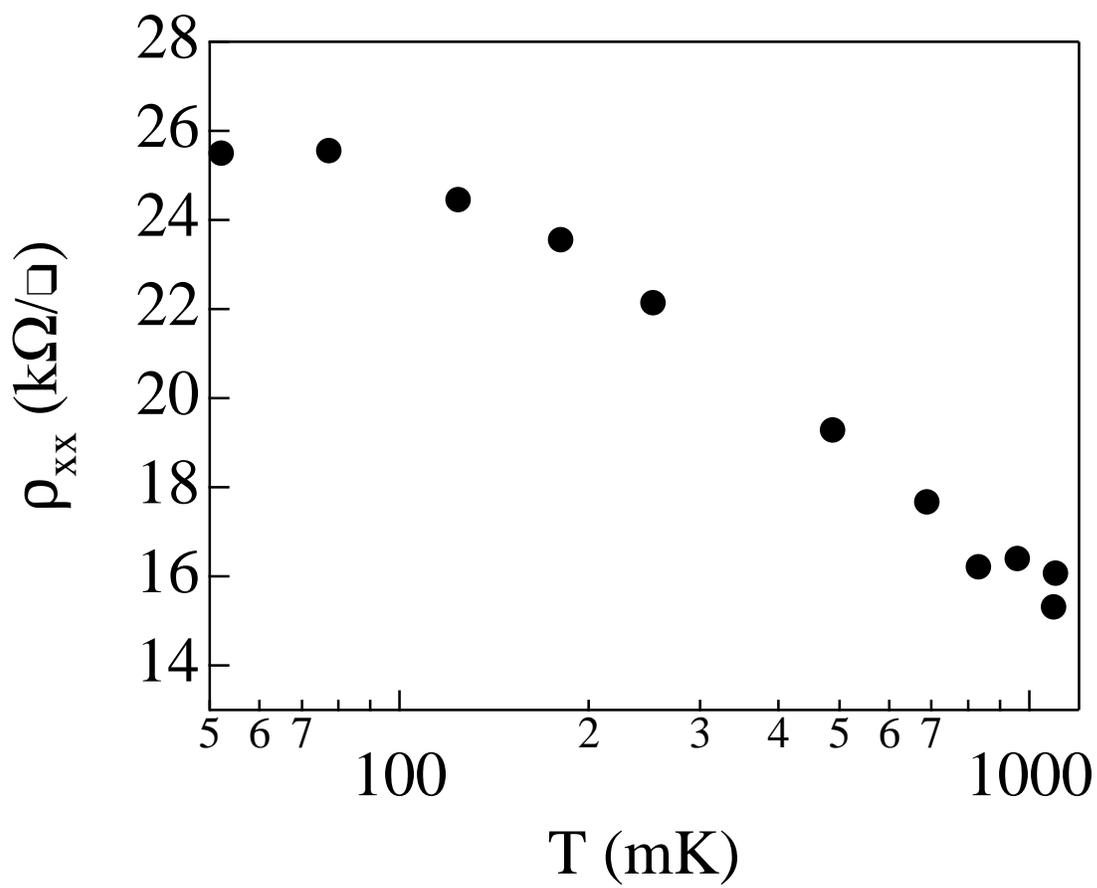
\begin{figure}
\caption{$\rho_{xxc}$ vs. $T$ for sample c60ab, with a log-$T$ scale.
}
\label{RxxTLog}
\end{figure}

So far, we reviewed our definition of the transition $B_c$ and discussed
the low-$T$ value of $\rho_{xxc}$. We now proceed to discuss the
$T$-dependence of $\rho_{xxc}$ at higher $T$'s. In Fig. \ref{RxxT} we plot
$\rho_{xxc}$ vs. $T$ for several samples. Depending on the sample and the
$T$ range at hand, different forms of behavior are observed. Common to
all samples is a certain range at the lowest $T$ where $\rho_{xxc}$ is
$T$ independent, as expected at the critical point of the transition.
This value of $\rho_{xxc}$ is the value plotted in Fig.
\ref{RxxcBc}. As $T$ is increased, systematic deviations are observed in most
samples. The $T$ where these deviations appear is sample dependent, as is
the trend which they take. In \ref{RxxTLog} we plot $\rho_{xxc}$ vs. $T$ on a
log scale which shows that for this particular sample, where we have a
relatively wide range of data, a logarithmic $T$ dependence of
$\rho_{xxc}$ is a reasonable description of the data. This dependence is
similar to that observed for 2D disordered metals at low $T$ and $B=0$.
It is not clear whether the various mechanisms that lead to the
logarithmic $T$-dependence at $B=0$ are applicable for the high-$B$ 
case.

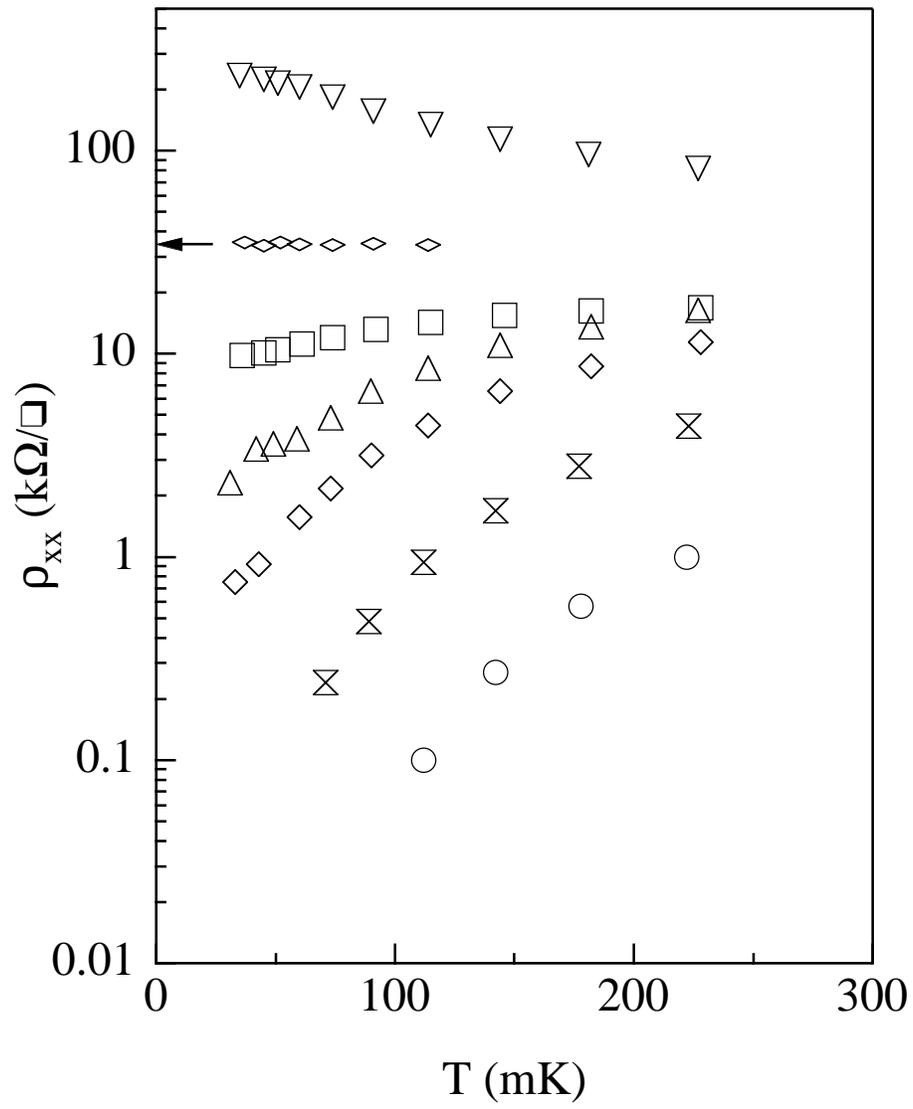
\begin{figure}
\caption{Disorder-induced QH-insulator transition. Densities are, 
from top to bottom, 0.7, 
1.05, 1.3, 1.5, 1.7, 2.1, 2.9 $10^{10}$ cm$^{-2}$.
}
\label{DisRxxT}
\end{figure}

In the limit of strong disorder, the QH ceases to exist and is replaced
with an insulating behavior. It is a reasonable expectation, that if
one could vary the effective disorder over a wide enough range,
a transition from a QH state to an insulator will be observed.
This expectation was verified in experiments\cite{jiang,wang,hughes}. In fact, the
disorder-induced transition was shown to be remarkably similar to the
$B$-field-induced one as far as its critical properties are 
concerned\cite{Wong95}. To
vary the effective
disorder the experimentalist usually employs a metallic gate deposited
near the 2D electrons. By biasing the gate with respect to the
electron system $n$ can be varied continuously, resulting in an effective
disorder change via the dependence of the impurities potential-strength
on the screening effectiveness of the electrons which, in turn, depends
on $n$.  In Fig. \ref{DisRxxT} we plot $\rho_{xxc}$ vs. $T$ in the vicinity of a
disorder-induced $\nu=1$ to insulator transition. The different sets of
data correspond, in this figure, to different gate-voltage bias and
therefore to different disorder. The qualitative similarity to the
$B$ induced transition is clear. In addition we note that $\rho_{xxc}$
for this transition is **24k$\Omega$, again close to $h/e^2$. The $T$ range
of our study in this case is not sufficient to detect the high-$T$
deviations of $\rho_{xxc}$.

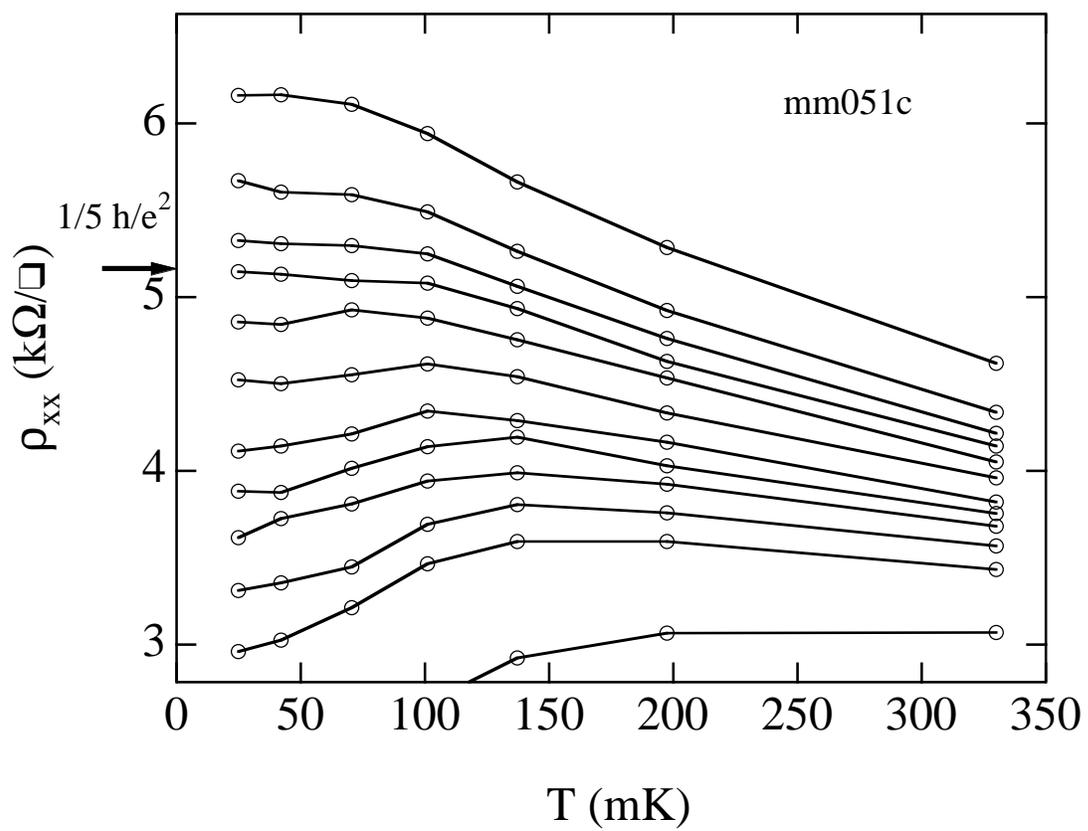
\begin{figure}
\caption{$\rho_{xx}$ vs. $T$ near a QH-QH transition. For this 
transition the predicted critical $\rho_{xx}$ is $1/5~h/e^{2}$.
}
\label{QHQH}
\end{figure}

Finally we remark on the behavior at the critical point of QH-QH
plateau-to-plateau transitions. As we demonstrated in a recent 
paper\cite{Shahar:QHtrans},
a direct and clear relation exist between these transitions: It is
possible to map the QH-to-insulator transition to a QH-QH transition by
considering the former as a QH-insulator transition occurring at the top
LL in the presence of an inert (full) bottom LL. In Ref. \cite{Shahar:QHtrans}
we found that the transition point, when properly identified, is at a 
value which is close (within 20\%) to the theoretically predicted 
value of $1/5 ~h/e^{2}$ (See Fig \ref{QHQH}). 
Many other samples exhibit QH-QH with $\rho_{xxc}$ much 
smaller than expected. We are uncertain why the QH-insulator 
transition yields a more consistent critical behavior than the QH-QH 
case. 

\section{Numerical study of Thouless conductance}
\label{theory}

In this section we present results of numerical studies of the Thouless
conductance at the critical point, using a non-interacting electron model
on a square lattice, described by the following Hamiltonian:
\begin{equation}
H=\sum_{m,n}\{-(c^\dagger_{m+1,n}c_{m,n}+c^\dagger_{m,n+1}e^{i2\pi \alpha m}
c_{m,n}+{\rm H.c.})
+\epsilon_{m,n}c^\dagger_{m,n}c_{m,n}\},
\label{hamilt}
\end{equation}
where the integers $m$ and $n$ are the $x$ and $y$ coordinates of the
lattice site in terms of lattice constant,
$c_{m,n}$ is the fermion operator on that site, 
H.c. stands for Hermitian conjugate, $\alpha$ is the amount of
magnetic flux per plaquette
in units of the flux quantum $hc/e$,
and $\epsilon$ is the random onsite potential.
We will present data mostly for uncorrelated random potential 
(i.e., no correlation
between $\epsilon$'s on different lattice sites), with $\epsilon$
ranging uniformly from $-W$ to $W$. Random potential
with some short range correlation will also be studied. 
The Landau gauge ${\bf A}=(0, Bx, 0)$ is used in Eq. (\ref{hamilt}). 
In this work we study finite size systems of square geometry, with linear
size $L$, for $L$ ranging from 18 to 50. 
We impose periodic boundary condition (PB) along the $\hat{x}$ direction:
$\Psi(k+L\hat{x})=e^{i\phi_1}\Psi(k)$, and periodic or antiperiodic
boundary condition (APB) along the $\hat{y}$ direction:
$\Psi(k+L\hat{y})=\pm \Psi(k)$.
We diagonalize the Hamiltonian (\ref{hamilt}) numerically to obtain the
single-electron spectrum for both the periodic boundary condition ($E^n_p$),
and antiperiodic boundary condition ($E^n_{ap}$) along $\hat{y}$ direction, 
while keeping the boundary condition along $\hat{x}$ to be periodic. Here
$n$ is the index for a specific eigenstate.

The Thouless conductance\cite{thouless} at Fermi energy $E$ is defined as
\begin{equation}
g_T(E)={\langle\delta E\rangle\over \Delta E},
\end{equation}
where $\Delta E=1/[L^2D(E)]$ is the average level spacing at $E$, determined by
the disorder averaged density of states (DOS) per site $D(E)$, 
and $\langle\delta E\rangle$ is the average of the
absolute value of the difference between 
$E^n_p$ and $E^n_{ap}$, also at energy $E$.\cite{note} 

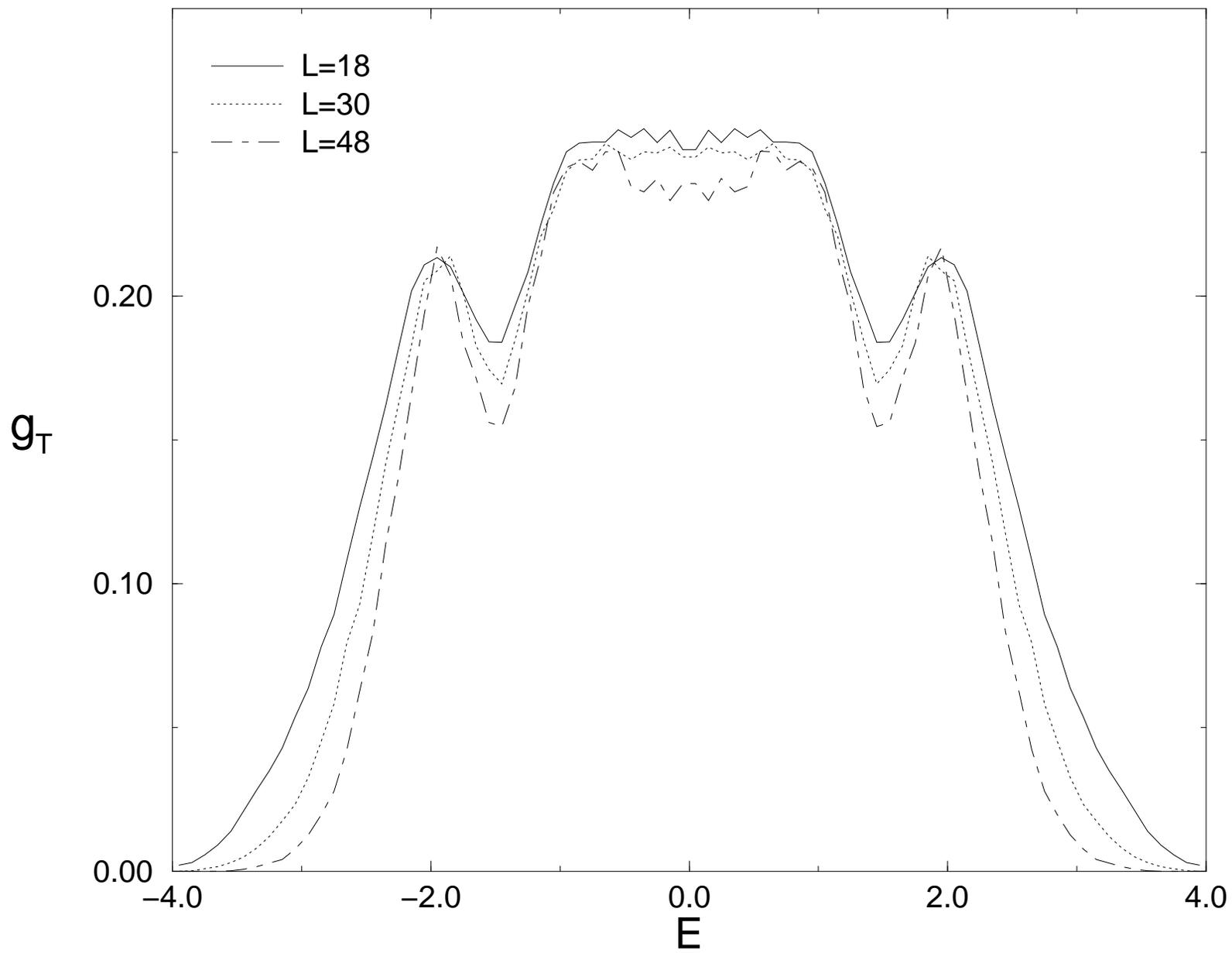
\begin{figure}
\caption{Thouless conductance $g_T$ as a function of energy for $\alpha=1/3$,
$W=2.5$, at different system sizes.
}
\label{size}
\end{figure}

In Fig. \ref{size} we show the Thouless conductance ($g_T$) for systems with
$\alpha=1/3$, $W=2.5$ (uncorrelated potential), and $L$ ranging from 18 to 48.
We find except for 4 special energies, $g_T$ {\em decreases} as $L$ increases;
while for $E=\pm E_c^1\approx\pm 2.0$ and $E=\pm E_c^2\approx\pm 1.1$, 
$g_T$ peaks, and appears to be essentially 
{\em independent} of $L$.\cite{indnote}  
The physics of such behavior may be understood in the following way.
\cite{tan,liu,yang,xie,sheng}
In the absence of random potential we have three Landau subbands, and the
Hall conductance (in unit of $e^2/h$) for each subband is 1 for the two side
bands and $-2$ for the central band. As random potential is turned on, 
most states get localized, but there will be one critical energy in each
side band ($\pm E_c^1$) and two critical energies ($\pm E_c^2$) 
in the central band, 
at which states 
are delocalized. The Hall conductance carried by the extend states is
1 for $\pm E_c^1$, and $-1$ for $\pm E_c^2$. For energies away from these
critical energies states are localized, therefore $g_T$ decreases as system
size $L$ increases, and goes to zero in the thermodynamic limit; 
at these critical energies states are delocalized, and $g_T$ approaches a 
finite number in the thermodynamic limit; for large enough system size $L$, 
$g_T$ is essentially independent
$L$. It is clear from the plot that $g_T$ has essentially reached its
asymptotic value for $L\ge 18$; the size dependence of $g_T$ at smaller sizes 
will be discussed later. 
$E_c^1$ and $E_c^2$ move together as $W$ increases, 
and at $W=W_c\approx 2.9$ they merge together and kill each other, and all
states become localized.\cite{yang}

In the following we will focus on the value of $g_T$ at the critical 
energies. In Fig. \ref{unc}a we show the size-independent 
peak value of $g_T$ at $E=-E_c^1$, for different randomness strength
$W$'s, and $L=30$ (which we believe to be in the asymptotic regime already). 
The value of $g_T$ is the same at $E=E_c^1$, due to particle-hole
symmetry of the model. We find $g_T\approx 
0.21\pm 0.02$, {\em independent} of the
value of $W$. In Fig. \ref{unc}b we present the
peak value of $g_T$ in the lowest Landau subband for $\alpha=1/5$ and
$\alpha=1/7$, at different $W$'s, and system sizes $L=25$ and $L=21$
respectively. We have checked that for these sizes the peak value of $g_T$ 
of the lowest Landau subband is already at its asymptotic value.
Again we get the same value, within
error bars, even though we have different field strength and different
number of Landau subbands.

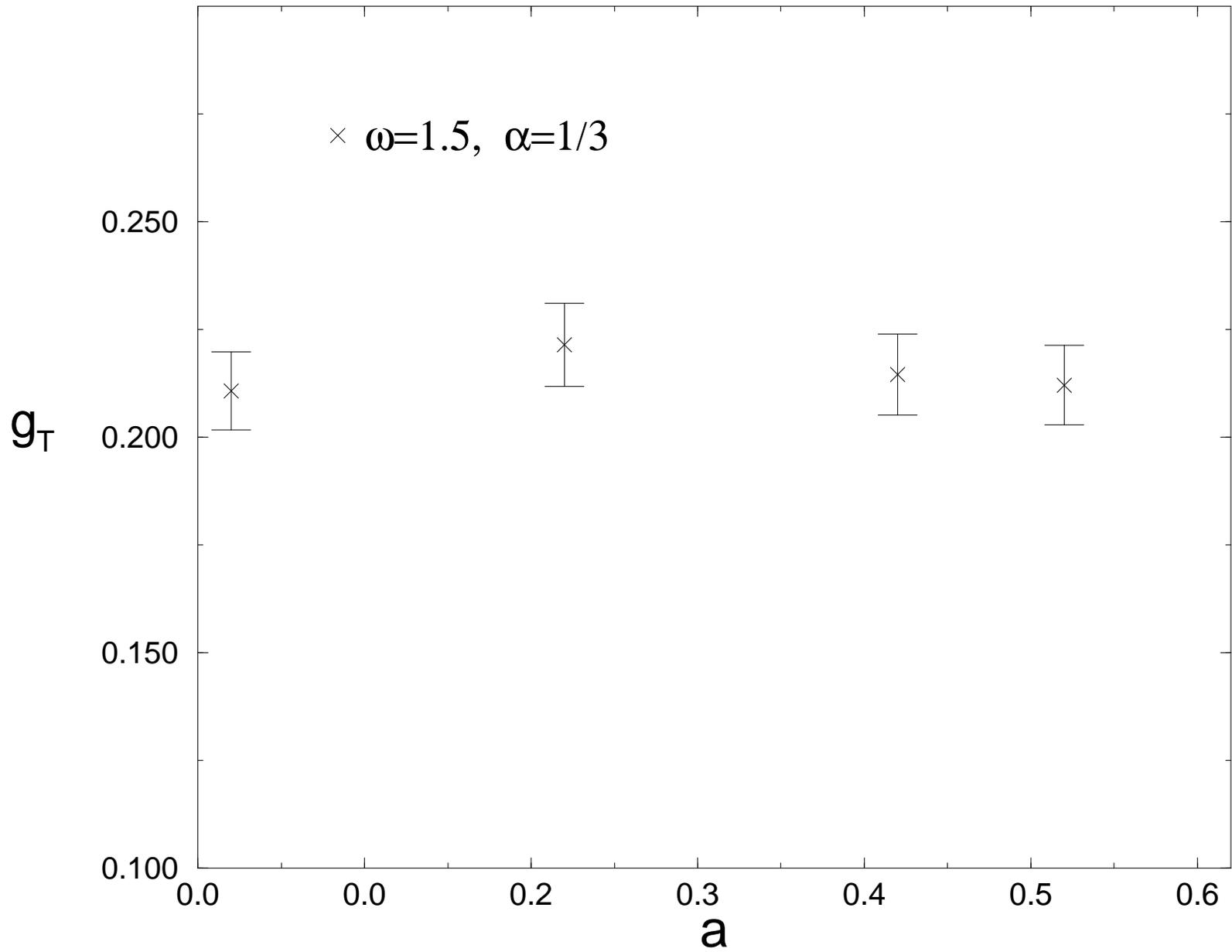
\begin{figure}
\caption{Peak value of the Thouless conductance $g_T$ in the lowest 
Landau subband for short-range correlated potential. $a$ is the strength of
short-range correlation.
}
\label{corr}
\end{figure}

So far we have only studied uncorrelated random potentials on the lattice,
which maps onto Gaussian
white noise potential in the continuum limit. In the
following we study random potentials with short-range correlations. We use
the following way to generate short-range correlation: numerically we
generate an uncorrelated random number $w_i$, uniformly distributed
from $-W$ to $W$, for each lattice site $i$. Instead of using $w_i$ as
the random potential $\epsilon_i$ as before, we take
\begin{equation}
\epsilon_i=w_i+a\sum_{\delta}w_{i+\delta},
\end{equation}
where the summation is over the four neighboring sites of $i$.
This way the potential of one site is correlated with its nearest and next
nearest neighbors, and the amount of correlation is determined by $a$.
In Fig. \ref{corr} we show the size-independent peak value of $g_T$
(again based on data with $L=30$)
at $E_c^1$,
for $\alpha=1/3$, $W=1.5$, at different $a$'s. We find 
within error bars it is independent of $a$, and takes the same value as
the uncorrelated potential ($a=0$).

Our data clearly indicates that $g_T\approx 0.21\pm 0.02$ 
is a universal number at the critical
energy of the lowest Landau subband of the square lattice, independent of
the strength of the randomness and magnetic field, as well as the type of
random potential (correlated or uncorrelated). 
Within error bars, the same universal number is also found in the lowest 
Landau level of the {\em continuum system} in Ref. \onlinecite{sorensen},
where the same definition of the Thouless conductance was used. 
We point out however that in our calculation, no projection to individual
subbands is made, and mixing between different Landau subbands (or levels)
(which is often important in real systems) is taken into account. 
We thus conclude that just like the conductivity tensor, the Thouless
conductance is a universal number at integer quantum Hall transition
in noninteracting electron models (either on a lattice or in the continuum).

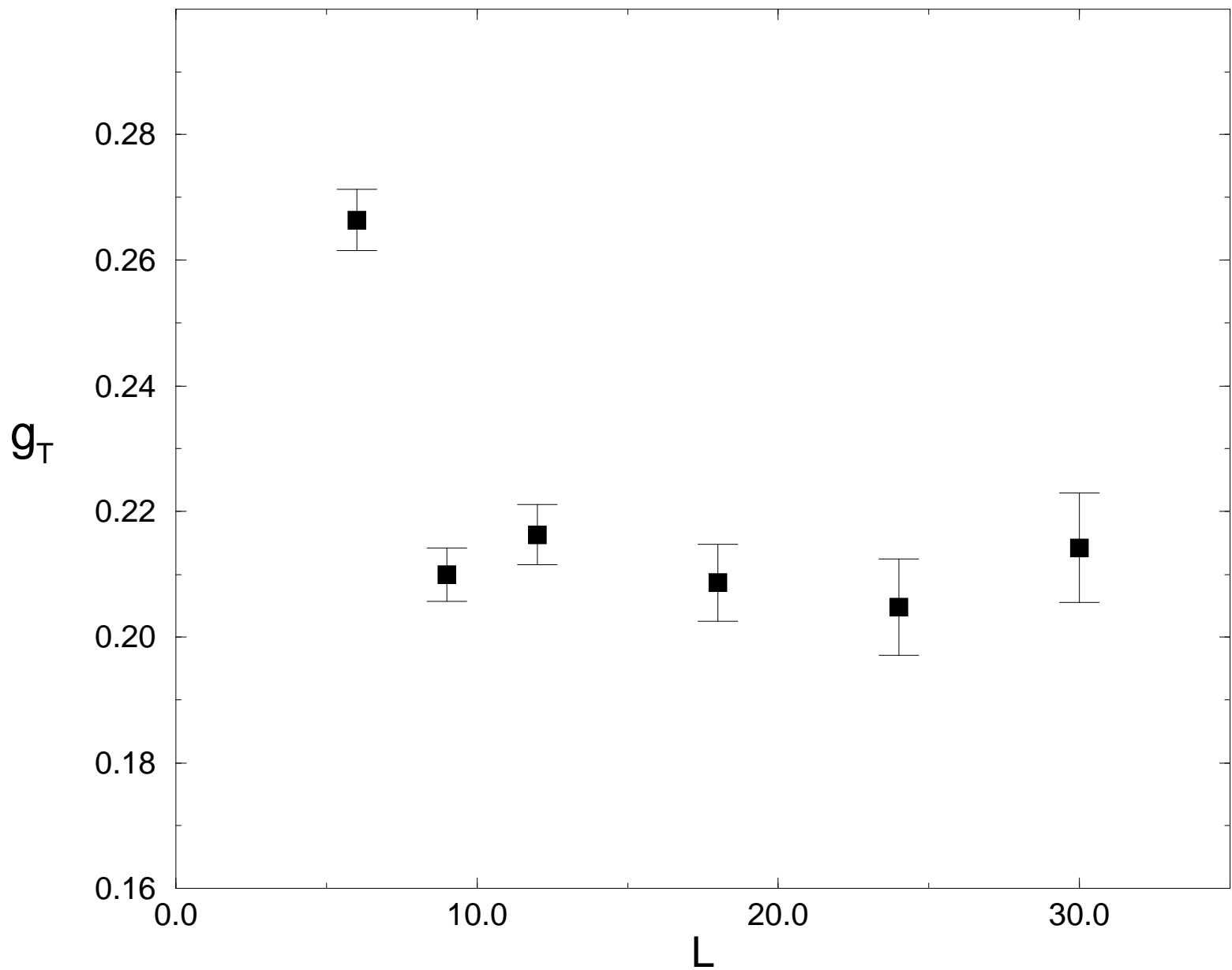
\begin{figure}
\caption{$L$ dependence of $g_T$ at the critical energy of the lowest 
Landau subband for $\alpha=1/3$.
}
\label{ld1}
\end{figure}

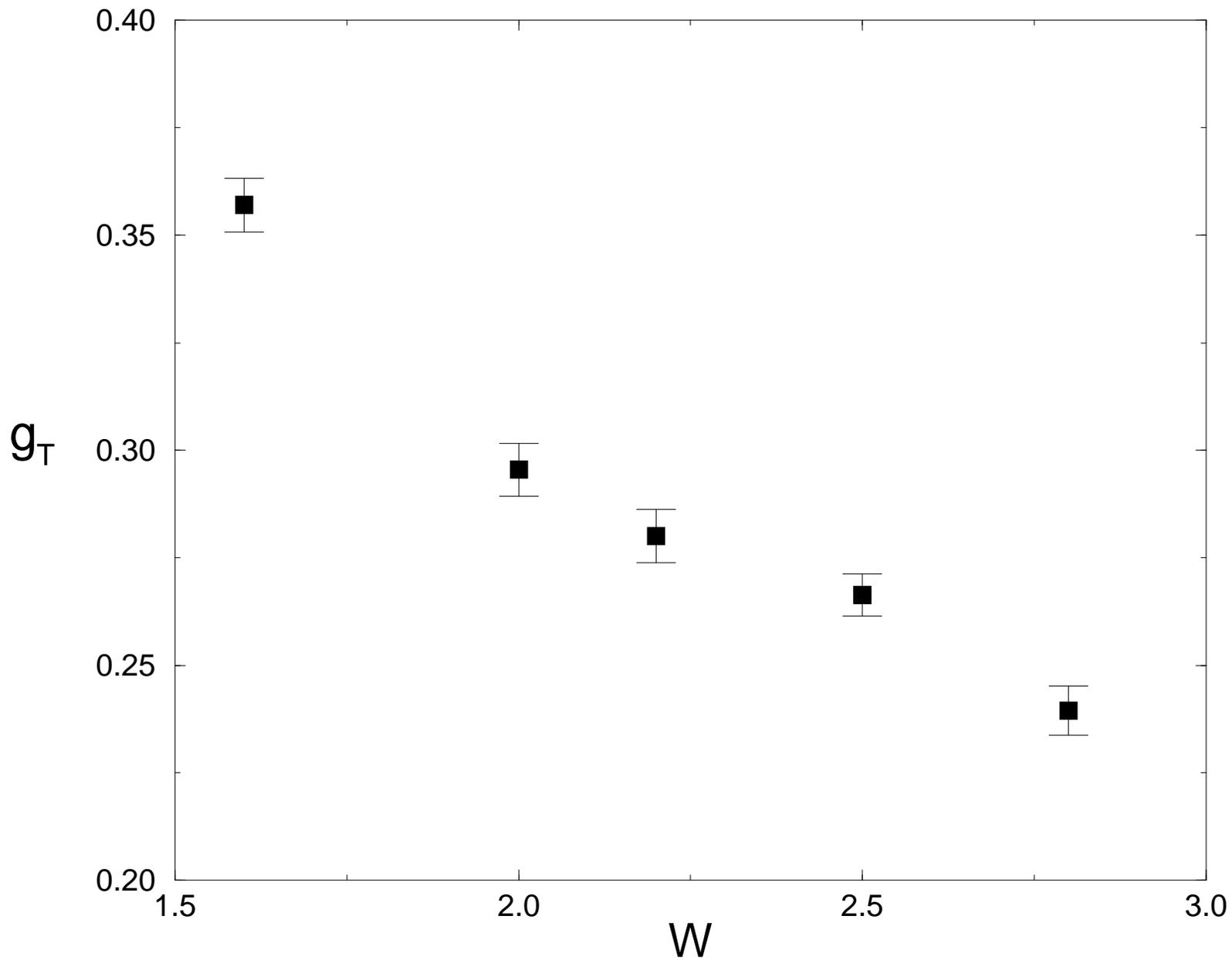
\begin{figure}
\caption{The peak value of $g_T$ (at $E=-E_c^1$)
in the lowest Landau subband for 
$\alpha=1/3$ and $L=6$ versus randomness strength $W$.
}
\label{wd}
\end{figure}

In principle, the truly universal value of $g_T$ at the critical energies
is reached in the thermodynamic
limit $L\rightarrow\infty$ only; there is always finite size correction of 
$g_T$ at finite $L$, and the correction should decrease as $L$ increases.
As discussed in section \ref{intro}, in a noninteracting system a finite
system size is equivalent to finite temperature in an infinite system. Since
real experiments are always done at finite temperatures, such finite size
effects are observable. Motivated by this we have also studied the size 
dependence of $g_T$ at the critical energies. 
In Fig. \ref{ld1} we plot the dependence of $g_T$ at $E=-E_c^1$ for 
$\alpha=1/3$ and $W=2.5$. We find, quite remarkably, that for $L$ as small
as 9, $g_T$ has essentially saturated at the asymptotic value, indicating 
that the finite size corrections of $g_T$ disappear
extremely fast as $L$ increases. The deviation of $g_T$ at $L=6$ from the
universal value is clearly due to finite size corrections; associated with 
that, we have also found strong dependence of the peak value of
$g_T$ at $L=6$ on the randomness strength $W$, as shown in Fig. \ref{wd}. 
We find the bigger $W$ is, the closer to the universal value $g_T$ becomes.
This is reasonable because the stronger randomness is more effective in 
localizing states away from critical energies, and therefore suppress 
finite size effects.
No such dependence on $W$,
however, 
is found for larger $L$ where $g_T$ has saturated at the universal value.

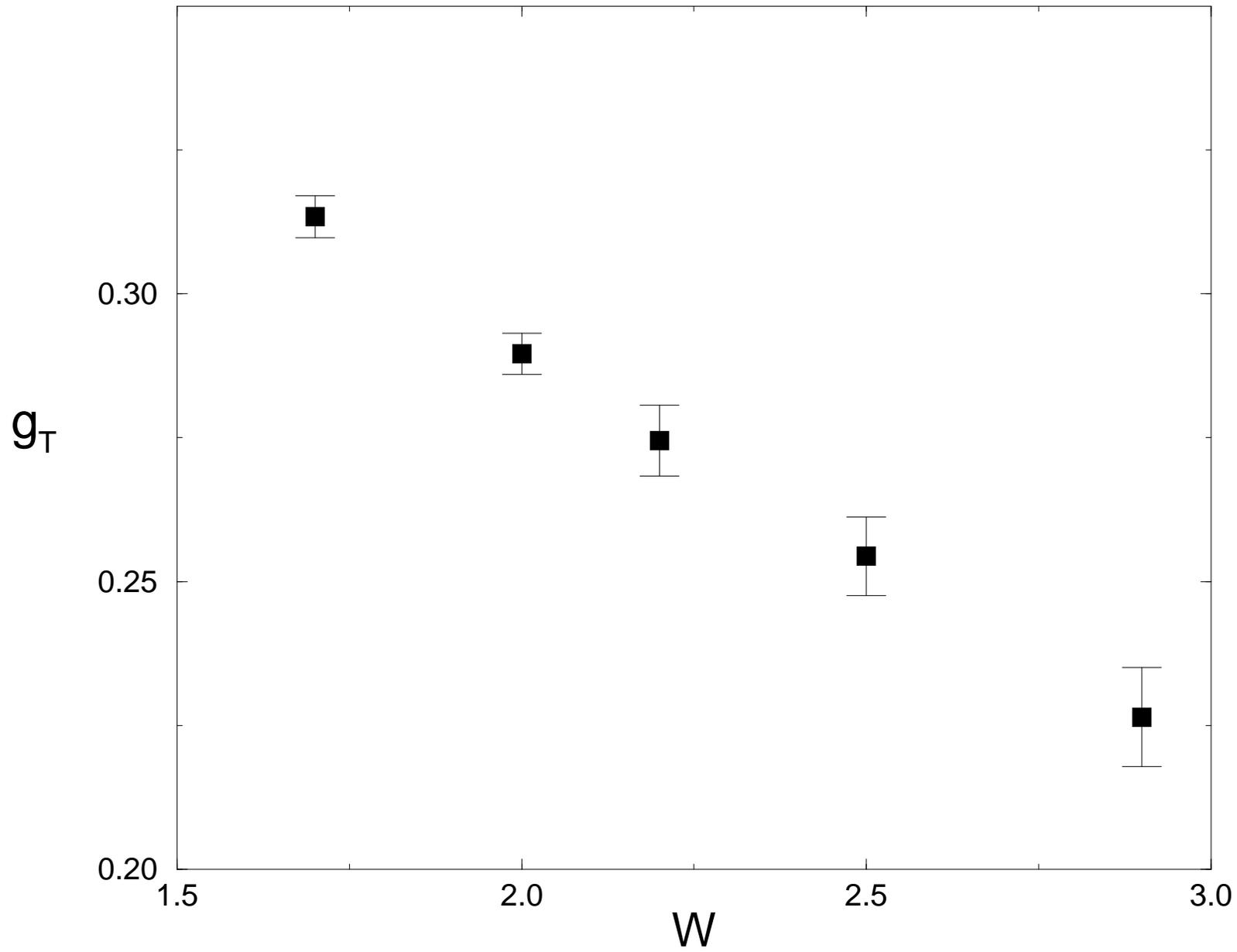
\begin{figure}
\caption{The peak value of $g_T$ (at $E=-E_c^2$)
in the central Landau subband for
$\alpha=1/3$ and $L=30$ versus randomness strength $W$ ($E_c^2$
depends on $W$).
}
\label{wd30}
\end{figure}

In the RG language, the finite size corrections to universal properties at the
critical point is due to the existence of irrelevant operators, whose strength
scales to zero in the thermodynamic limit under RG at the critical point, 
while they remain finite in finite size systems. It has been known\cite{bodo}
for some time now, based on numerical studies, 
that the length scale required for such irrelevant operators
to scale
away is quite small in the lowest Landau level, while in higher Landau
levels it becomes very large. The origin of this difference is not yet
fully understood. Our results in the lowest Landau subband is clearly 
consistent with this finding. Also consistent with this, we do see some 
weak size dependence of the peak value of $g_T$ in higher subbands (see
Fig. \ref{size}), suggesting the existence of finite size correction in the
size range of our numerical study. We have also found strong dependence of
the peak value of $g_T$ in higher subbands on $W$ for a given size as 
shown in Fig. \ref{wd30}. The dependence is very similar to that 
of the peak value of $g_T$ in the lowest subband with $L=6$, 
where we know finite
size corrections are present. Based on these we conclude that 
finite size corrections are quite important in higher Landau subbands within
the size range of the present study, and conjecture that
in the thermodynamic limit,
the peak value of $g_T$ will saturate at the same universal value as in the
lowest subband, provided that the critical energy carries Hall conductance
$\pm 1$.  
We will discuss the possible experimental  
consequences of these finite size effects in section \ref{summary}.

\section{Summary and Discussions}
\label{summary}

In this paper we have presented results of detailed experimental studies on 
the universality of the resistivity tensor at the quantum critical points
separating an integer quantum Hall phase
and the high magnetic field insulator, as well
as critical points separating different integer 
quantum Hall phases. Our results 
strongly suggest that the resistivity tensor is universal at the critical
points, and that
quantum Hall-insulator and quantum Hall-quantum Hall transitions are in the
same universality class. This is in agreement with previous experimental
studies, as well as general theoretical expectations.

We have also studied such transitions using a non-interacting electron model
on a lattice, and found that the Thouless conductance at the critical points
is universal. This is in agreement with our experimental findings, as well as
previous theoretical studies using different models and approaches.
It has been known for a long time that 
the Thouless conductance is closely related to the longitudinal
conductivity of the system; they are believed to be of the same order
of magnitude, no matter the system is in the insulating phase, metallic
phase, or at the critical point, and are roughly proportional to each other.
People have not succeeded, however, in
establishing an exact relation between these two quantities in general.
If both of them are universal at the
critical point, their ratio must also be universal at the critical point.
Since numerically the Thouless conductance is much easier to calculate
than the longitudinal conductivity, such a simple relation should be
very useful in future researches.
It will also be very interesting to
see if the same ratio relates these two quantities away from the
critical point.

We have also, probably for the first time in the literature, presented a 
detailed analysis on how the universal values of the resistivity tensor are
approached in the asymptotic low temperature regime, and the deviation 
from the universal values at higher temperatures. We find that the logitudinal
resistivity $\rho_{xx}$ at the critical point has some sizable, and 
apparently non-systematic deviation from the universal value at relatively
high temperatures; while the Hall resistivity $\rho_{xy}$ has essentially no
such deviation in the same temperature range. Using the relation betweem the
resistivity and conductivity tensors:
\begin{eqnarray}
\rho_{xx}={\sigma_{xx}\over \sigma_{xx}^2 +\sigma_{xy}^2},\\ 
\rho_{xy}={\sigma_{xy}\over \sigma_{xx}^2 +\sigma_{xy}^2},
\end{eqnarray}
we find the deviation of $\rho_{xx}$ and $\rho_{xy}$ from the universal
values are related to the deviation of $\sigma_{xx}$ and $\sigma_{xy}$
to the lowest order\cite{devnote} in the following way:
\begin{eqnarray}
\delta\rho_{xx}\approx{1\over \sigma_{xx}^2 +\sigma_{xy}^2}
\left[\delta\sigma_{xx}(1-{2\sigma_{xxc}^2\over \sigma_{xxc}^2 +\sigma_{xyc}^2})
-\delta\sigma_{xy}{2\sigma_{xyc}^2\over \sigma_{xxc}^2 +\sigma_{xyc}^2}
\right],\\
\delta\rho_{xy}\approx{1\over \sigma_{xx}^2 +\sigma_{xy}^2}
\left[\delta\sigma_{xy}(1-{2\sigma_{xyc}^2\over \sigma_{xxc}^2 +\sigma_{xyc}^2})
-\delta\sigma_{xx}{2\sigma_{xxc}^2\over \sigma_{xxc}^2 +\sigma_{xyc}^2}
\right].
\end{eqnarray}
Since at the integer quantum Hall-insulator transition
$\sigma_{xxc}=\sigma_{xxc}=0.5e^2/h$, we find
\begin{eqnarray}
\delta\rho_{xx}\approx -\delta\sigma_{xy}(2h^2/e^4),\\
\delta\rho_{xy}\approx -\delta\sigma_{xx}(2h^2/e^4).
\end{eqnarray}
Thus the deviation of {\em longitudinal} resistivity 
$\delta\rho_{xx}$ at finite $T$ is proportional to the deviation of {\em Hall}
resistivity $\delta\sigma_{xy}$, while the absence of deviation in {\em Hall}
resistivity implies the deviation of {\em longitudinal} conductivity 
$\delta\sigma_{xx}\approx 0$! In our numerical study on
Thouless conductance, we have already seen that the Thouless conductance, 
which is believed to be proportional to $\sigma_{xx}$, approaches the 
universal value at extremely small system sizes at the critical point; 
this naturally explains the
fact $\delta\sigma_{xx}\approx 0$ and hence the absence of deviation 
in $\rho_{xx}$ at finite $T$ at the critical point. 
In a noninteracting electron model, $\delta\sigma_{xy}$ at the critical point
due to finite size effects can only come from particle-hole (PH) asymmetry in
the corresponding Landau level (band). Both PH asymmetry in the underlying
potential, and mixing  
of different Landau levels (bands), can give rise to such asymmetry.
A detailed numerical study on the effects of PH asymmetry on 
$\delta\sigma_{xy}$ at the critical point in finite size systems will be 
presented elsewhere.

Quantum phase transitions is a fascinating subject, and we expect study of
quantum Hall transitions continues to be a fruitful field of research. 

\acknowledgements
This work was supported by NSF grant DMR-9400362.

\begin{figure}
\caption{Peak value of the Thouless conductance $g_T$ in the lowest 
Landau subband, for different magnetic field ($\alpha$)
and randomness ($W$) strength, with uncorrelated random potential.
}
\label{unc}
\end{figure}

\begin{references}
\bibitem{note} There are different ways to define the Thouless conductance.
For example, Ref. \onlinecite{macdonald} uses the geometric mean instead of
the ordinary average of the energy difference to determine $\langle\delta E
\rangle$. What we mean by the universality of Thouless conductance is for
a fixed definition, like the one used in this paper, the 
Thouless conductance at the critical point is independent of the details 
of the model.
\bibitem{indnote} The independence of $g_T$ on $L$ is quite clear for 
$E=\pm E_c^1$, while at $\pm E_c^2$ there appear to be some very weak $L$ 
dependence. Based on previous studies (see below), we know that there should
be two critical energies at $\pm E_c^2$. The weak size dependence of $g_T$ on 
$L$ will be discussed later.
\bibitem{devnote} Since the observed deviation at finite $T$
is typically of order one
percent of the asympototic value, expansion to the lowest order should be
sufficient.
\end{references}
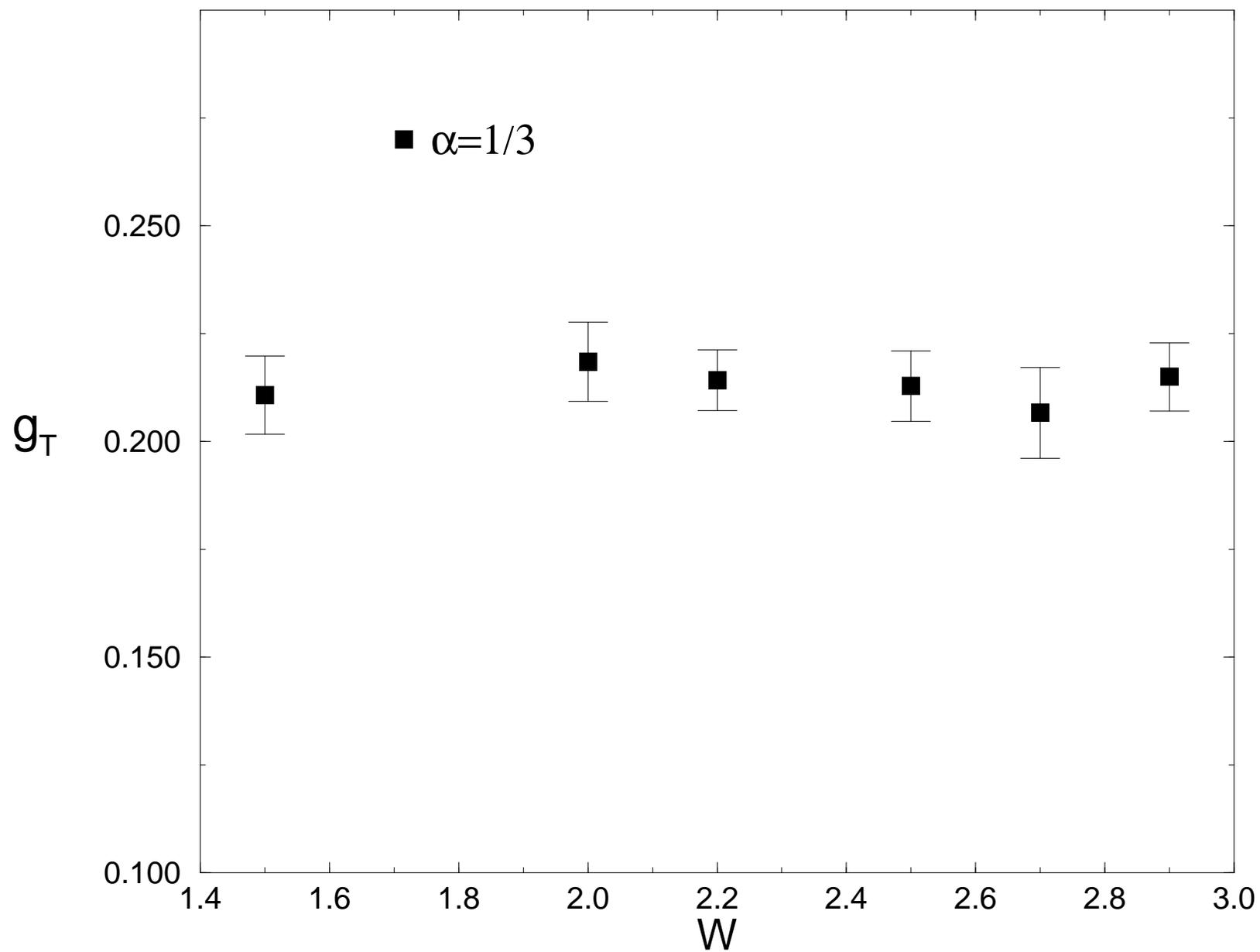
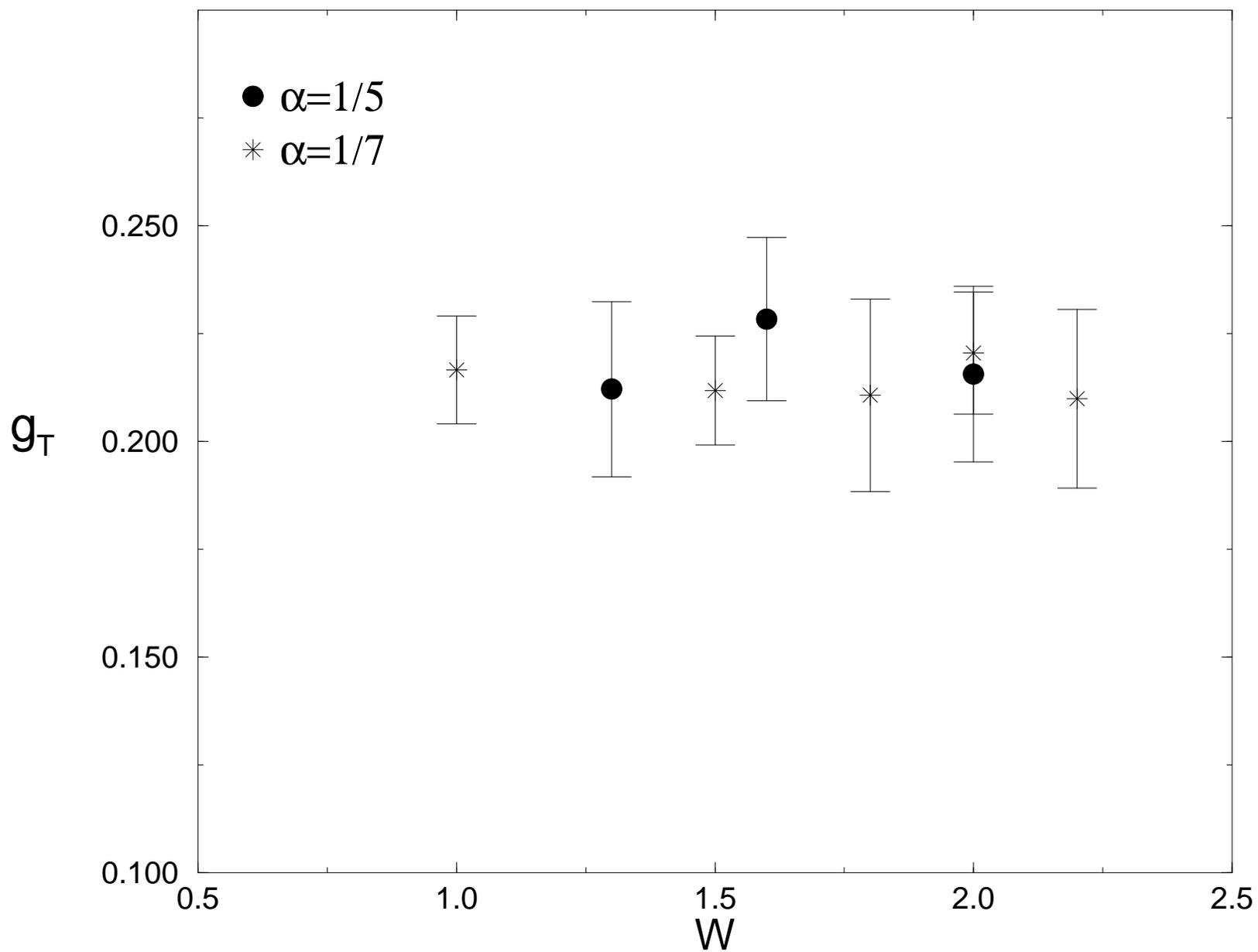
\end{document}